\documentclass[letterpaper, 10 pt, conference]{ieeeconf}  % Comment this line out if you need a4paper

\IEEEoverridecommandlockouts                              % 
\usepackage{graphicx} % for pdf, bitmapped graphics files
\usepackage{epsfig} % for postscript graphics files
\usepackage{soul,color} 
\usepackage{xcolor}
\usepackage{comment}
\usepackage{xspace}
\usepackage{mathtools}
\usepackage{booktabs}
\usepackage{multirow}
\usepackage{enumerate}
\usepackage{balance}
\usepackage{gensymb}
\usepackage{tabulary}
\usepackage{dblfloatfix}
\usepackage{layouts}
\usepackage{algorithm}
\usepackage{algpseudocode} % http://ctan.org/pkg/algorithmicx
\usepackage{cite}
\usepackage{bm}
\usepackage{hyperref}
\usepackage{svg}
\usepackage{cite}
\usepackage{caption}
\usepackage{subcaption}
\usepackage{colortbl}
\usepackage{overpic}
\usepackage{rotating}
\usepackage{afterpage}
\usepackage{float}
\usepackage{url}
\usepackage{wrapfig}
\usepackage{listings}
\usepackage{adjustbox}
\usepackage{mdwtab}
\usepackage{array}
\usepackage{setspace}
\usepackage{mathdefs}
\usepackage{sonar}
\usepackage{units}
\usepackage{makecell}
\usepackage[many]{tcolorbox}

\newcommand{\spatialpoint}{\ensuremath{\mathbf{x}}}

\newcolumntype{?}{!{\vrule width 2pt}}
\definecolor{lightgreen}{rgb}{0.8,1,0.8}

\usepackage{caption}
\graphicspath{{../figs/}}

\makeatletter
\newcount\my@repeat@count
\newcommand{\myrepeat}[2]{%
  \begingroup
  \my@repeat@count=\z@
  \@whilenum\my@repeat@count<#1\do{#2\advance\my@repeat@count\@ne}%
  \endgroup
}

\title{\LARGE \bf
Acoustic Neural 3D Reconstruction Under Pose Drift
}

\author{{\fontsize{10}{12}\selectfont Tianxiang Lin$^{*1}$, Mohamad Qadri$^{*1}$, Kevin Zhang$^2$, Adithya Pediredla$^3$, Christopher A. Metzler$^2$, Michael Kaess$^1$} % <-this % stops a space
\thanks{* Indicates equal contribution}% <-this % stops a space
\thanks{$^{1}$T. Lin, M. Qadri, M. Kaess are with the Robotics Institute, Carnegie Mellon University, Pittsburgh, PA, USA.
        {\tt\small \{tianxian, mqadri, kaess\}@cs.cmu.edu}}%
\thanks{$^{2}$K. Zhang, C. Metzler are with the Department of Computer Science at the University of Maryland, College Park.
         {\tt\small \{kzhang24, metzler\}@umd.edu}}
\thanks{$^{3}$A. Pediredla is with the Computer Science Department at Dartmouth College.
    {\tt\small adithya.k.pediredla@dartmouth.edu}}
}

\begin{document}

\maketitle
\thispagestyle{empty}
\pagestyle{empty}

%%%%%%%%%%%%%%%%%%%%%%%%%%%%%%%%%%%%%%%%%%%%%%%%%%%%%%%%%%%%%%%%%%%%%%%%%%%%%%%%
\begin{abstract}
We consider the problem of optimizing neural implicit surfaces for 3D reconstruction using acoustic images collected with drifting sensor poses. The accuracy of current state-of-the-art 3D acoustic modeling algorithms is highly dependent on accurate pose estimation; small errors in sensor pose can lead to severe reconstruction artifacts. In this paper, we propose an algorithm that jointly optimizes the neural scene representation and sonar poses. Our algorithm does so by parameterizing the 6DoF poses as learnable parameters and backpropagating gradients through the neural renderer and implicit representation. We validated our algorithm on both real and simulated datasets. It produces high-fidelity 3D reconstructions even under significant pose drift. 
\end{abstract}

%%%%%%%%%%%%%%%%%%%%%%%%%%%%%%%%%%%%%%%%%%%%%%%%%%%%%%%%%%%%%%%%%%%%%%%%%%%%%%%%
\section{Introduction}
\noindent  Autonomous Underwater Vehicles (AUVs) often carry imaging sonar, also known as forward-looking sonar (FLS). 
Unlike an optical camera, an imaging sonar is able to capture long-range information in turbid conditions. 
Thus, because of its robustness, FLS has been integrated into many underwater applications, such as underwater inspection, construction, ecology, archaeology, and surveillance \cite{wang2019underwater,negahdaripour2018application,albiez2015flatfish,Lin23icra}.

Imaging sonar captures 2D measurements by emitting acoustic pulses and measuring the intensity and arrival time of reflections from 3D structures. Using beamforming and time-of-flight techniques, an imaging sonar can recover azimuth and range information. However, a key limitation is that it does not provide direct elevation measurements, making it inherently ill-suited for applications that require 3D information.

To reconstruct 3D structures, prior methods \cite{qadri23icra,aykin2016three,feng2024differentiable} tried to mitigate this limitation by integrating multiple imaging sonar measurements taken from known and precise poses. However, these approaches rely heavily on accurate poses and any errors in these poses will significantly degrade reconstruction quality.

To mitigate the impact of pose errors, we propose a technique that jointly optimizes the 3D structure and the sonar poses. By incorporating pose optimization directly into the reconstruction process, our method improves robustness to pose drift and sensor noise, enabling more reliable 3D reconstruction from imaging sonar.
Our contributions are as follows: 

\begin{itemize}
    \item A framework for recovering 3D structure from acoustic sonar measurements while simultaneously optimizing sensor poses.
    \item A qualitative analysis of the convergence properties of the optimized pose solution set.
    \item Evaluation on both real and simulated datasets containing objects with varied geometries. 
\end{itemize}

%%%%%%%%%%%%%%%%%%%%%%%%%%%%%%%%%%%%%%%%%%%%%%%%%%%%%%%%%%%%%%%%%%%%%%%%%%%%%%%%
\section{Related Work}
\subsection{3D Reconstruction From Imaging Sonar}
\noindent  Prior works introduce a variety of techniques for reconstructing 3D structures from imaging sonar measurements, such as ICP-based alignment \cite{teixeira2016underwater}, space carving \cite{aykin20153,aykin2016three}, solving constraint equations \cite{westman2020theory}, generative sensor modeling \cite{westman2019wide}, graph-based processing \cite{wang20183d,wang2019three}, convex optimization \cite{westman2020volumetric}, and supervised learning \cite{debortoli2019elevatenet,wang2021elevation,arnold2022spatial}.

More recently, differentiable rendering-based techniques have achieved state-of-the-art performance in the reconstruction of 3D structures from sonar imagery.  Qadri et al.~\cite{qadri23icra} propose combining a neural surface representation with a novel differentiable acoustic volume rendering process to recover 3D surfaces from imaging sonar data. In an extension \cite{10.1145/3641519.3657446} they combine optical camera information with sonar imagery to achieve improved 3D reconstruction performance in the small baseline setting. Reed et al.~\cite{10.1145/3592141} recover 3D volumes from synthetic aperture sonar measurements using neural rendering. Xie et al.~\cite{10631294, 10832111} use neural rendering to perform bathymetry from imaging sonar and sidescan sonar, respectively. Qu et al.~\cite{10685550} derive a novel forward splatting process for Gaussian Splatting and use it to recover 3D structures from sonar imagery. 

All of these methods rely critically upon reliable estimates of the poses at which the used sonar imagery is captured, but none as of yet concentrate on recovering 3D scenes from noisy poses. 

\subsection{Pose Optimization}
\noindent  Conventional underwater simultaneous localization and mapping (SLAM) methods have been widely explored and applied to solve underwater navigation problems. These methods rely on well-studied algorithms from state estimation \cite{kaess2012isam2, dellaert2017factor, qadri2022incopt, qadri2024learning, qadri2023learning}. Shin et al. \cite{Shin15Ocean} explore a pairwise bundle adjustment method by exploiting spatial constraints using KAZE features \cite{KAZE} between paired sonar images, which are refined by random sample consensus (RANSAC). Acoustic Structure-from-Motion (ASFM) algorithm to recover poses from multiple sonar images and drifting odometry \cite{Huang15iros}. Westman et al. \cite{westman2019degeneracy} improve the performance of multiview pose optimization by adding two-view sonar constraints during loop closure detections. Loi et al. \cite{Loi24Ocean} introduce submap registration for point cloud alignment or feature matching. Xu et al. \cite{Xu24ICRA} propose a direct imaging sonar odometry system that minimizes the aggregated two-view reprojection errors of sonar pixels with high-intensity gradients. These prior works fail to recover poses from the drifting odometry when sonar images provide limited features. At the same time, sonar intensity values are determined by multiple factors such as the orientation of the sonar with respect to the object, objects' material, etc.  Sonar images with significant speckle noise and multi-path effects can lead to the failure of classical underwater SLAM due to errors in feature matching.

Improving neural rendering reconstructions by simultaneously optimizing the reconstruction and poses is an active area of research. Wang et al. \cite{wang2022nerfneuralradiancefields} propose using an axis-angle and translation parameterization of camera poses and optimizing them with a neural radiance field simultaneously. Lin et al. \cite{Lin2021BARFBN} use a coarse-to-fine optimization strategy along with a neural radiance field pipeline to recover image poses from a collection of images. Bian et al. \cite{Bian2022NoPeNeRFON} exploit monocular depth priors to improve joint estimation of poses and neural radiance fields from images. Chng et al. \cite{Chng2022GaussianAN} use Gaussian activations to improve joint estimation of poses and neural radiance fields. Park et al. \cite{Park2023CamPCP} precondition camera parameters leading to improved reconstruction during the joint estimation of poses and neural radiance fields. All these prior works concentrate on 3D reconstruction from optical images with noisy camera poses. In this work, we focus on 3D reconstruction from acoustic images with noisy sonar poses. 

\begin{figure*} [th]
    \centering
    \includegraphics[width=0.83\paperwidth]{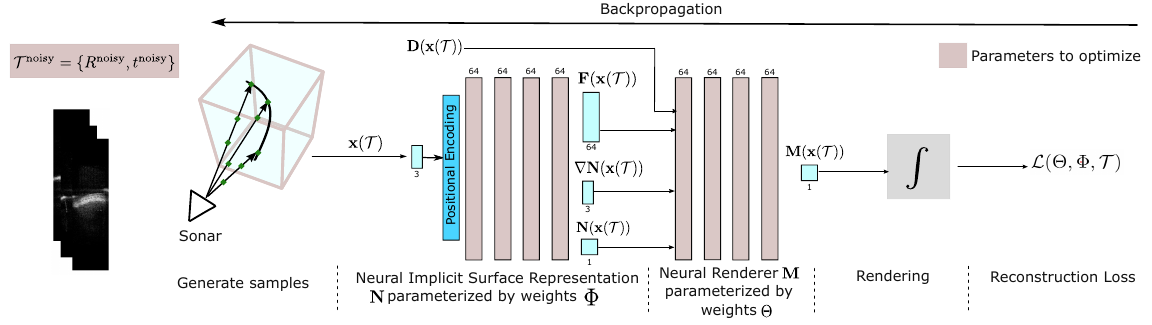}
    \caption{The pipeline of our proposed method. Our approach jointly optimizes sonar neural implicit surface networks and pose parameters by minimizing the total reconstruction loss. It takes 3D samples and viewing directions—both dependent on pose estimates—as inputs and outputs the signed distance function (SDF) $\mathbf{N}$ and outgoing acoustic radiance $\mathbf{M}$ for sonar image rendering. This pipeline enables training with sonar images and odometry that may be subject to drift.}
    \label{fig:architecture}
\end{figure*}

%%%%%%%%%%%%%%%%%%%%%%%%%%%%%%%%%%%%%%%%%%%%%%%%%%%%%%%%%%%%%%%%%%%%%%%%%%%%%%%%
\section{Background on Acoustic Neural Rendering}
\noindent  Given a training set of posed acoustic sonar images, the objective is to retrieve an accurate 3D reconstruction of an object of interest. We review \textit{NeuSIS}, a neural rendering method presented by Qadri et al.~\cite{qadri23icra} upon which our work is based. The technique allows for state-of-the-art performance for acoustic rendering by leveraging neural implicit surfaces and introduces a novel volumetric rendering equation. 

\subsection{Image Formation Model}
\noindent  Imaging sonars emit acoustic pulses and measure the intensity of the reflected signals to form a 2D acoustic image. While range and azimuth are resolved by the sensor, elevation remains ambiguous: The intensity of each pixel in the image is proportional to the cumulative sum of acoustic energy reflected by objects intersected by the acoustic arc at a specific range and azimuth. Hence, Qadri et al. \cite{qadri23icra} use the following image formation model where, for each pixel, the intensity $I_p$ at pixel $p = (r_i, \theta_i)$ is modeled as the integral of the reflected acoustic energy along each ray over the acoustic arc: 
\begin{align}
& I_p =  \int_{\phi_\text{min}}^{\phi_\text{max}} \int_{r_i-\epsilon}^{r_i+\epsilon} \frac{\textit{E}_e}{r} T(r, \theta_i, \phi) \sigma(r, \theta_i, \phi) \text{d}r\text{d}\phi.
\label{renderingloss}
\end{align}
where $\phi_\text{min}, \phi_\text{max}$ are the minimum and maximum elevation angles,  $E_e$ is the acoustic energy emitted by the sonar. $T=e^{-\int_0^{r_i} \sigma(r', \theta_i, \phi_i ) \text{d}r'}$ is the transmittance term, and $\sigma$ is the particle density.

\subsection{Neural Representation}
\noindent  Similar to Yariv et al. \cite{yariv2020multiview}, the object is represented as Signed Distance Function (SDF), $\mathbf{N}(\mathbf{x})$, which outputs the distance of each 3D point $\mathbf{x}=(X,Y,Z)$ to the nearest surface. A separate network, $\mathbf{M}$, computes $\mathbf{M}$($\mathbf{x})$,  the outgoing acoustic radiance at each spatial coordinate $\mathbf{x}$ which is then used to approximate the intensity of each pixel $\hat{I}_p$. The accuracy of this intensity estimate is correlated with the accuracy of the SDF representation: if the SDF is close to the ground-truth object then pixel $p$ of the $i$th training image $\hat{I}_p^i\rightarrow I_p^i$. 

Note that in this work, we assume noisy pose estimates. Hence, the approximated pixel intensity will depend on both the SDF representation as well as on our current estimate of the sonar poses.

\subsection{Approximation of the Image Formation Integral}
\noindent  Eq.~\ref{renderingloss} is discretized and approximated by sampling 3D points along both acoustic rays and arcs. 
\begin{align}
&    \hat{I}_p = \sum_{\mathbf{x} \in \mathcal{A}_p}
\frac{1}{r(\mathbf{x})} T[\mathbf{x}] \alpha[\mathbf{x}] \mathbf{M}(\mathbf{x})
\label{discreteImageFormationModelSonar}
\end{align}
where $\mathcal{A}_{p}$ is the set of sampled points along the acoustic arc at pixel $p$ = $(r_i, \theta_i$) and $ \mathbf{M}$ is the predicted radiance at $\mathbf{x}$. The computations of the discrete transmittance $T[\mathbf{x}]$ and opacity $\alpha[\mathbf{x}]$ terms require additionally sampling along acoustic rays. For any such spatial sample $\mathbf{x}_s$ on the  acoustic ray, the discrete opacity at $\mathbf{x}_s$ can be approximated as
 \begin{align}
    \alpha[\mathbf{x}_s] = \max\left(\frac{\mathbf{\Phi}_s(\mathbf{N}(\mathbf{x}_{s}))-\mathbf{\Phi}_s(\mathbf{N}(\mathbf{x}_{s+1}))}{\mathbf{\Phi}_s(\mathbf{N}(\mathbf{x}_{s}))}, 0\right),
    \label{discreteopacity}
\end{align} 
where $\mathbf{\Phi}_s(x) = (1+e^{-sx})^{-1}$ is the sigmoid function and $s$ is a trainable parameter while the discrete transmittance is modeled as 
\begin{align}
    T[\mathbf{x_s}] = \prod_{\mathbf{x_r} \;|\; r < s} (1-  \alpha[\mathbf{x_r}]).
\end{align}

%%%%%%%%%%%%%%%%%%%%%%%%%%%%%%%%%%%%%%%%%%%%%%%%%%%%%%%%%%%%%%%%%%%%%%%%%%%%%%%%
\section{Method}

\subsection{Loss Function}
\noindent We define the following set of trainable parameters: \noindent $\Theta$ are the weights of the SDF network $M$, $\Phi$ are the weights of the neural renderer $N$,  and $\mathcal{T} = \{T_i\}$ which parametrize the learnable sonar poses. Any intensity value computed via Eq.~\ref{discreteImageFormationModelSonar} is a function of $\Theta$, $\Phi$, and $\mathcal{T}$ - in other words, we can define $\hat{I}_p(\Theta, \Phi, \mathcal{T}) $ as the predicted intensity of the  $p$th  pixel of a training image and express our rendering loss function in terms of these three sets of parameters.

Our loss function is composed of three terms: 
The intensity loss  
\begin{align}
\mathcal{L}_{\text{int}} \equiv \frac{1}{|\mathcal{P}|} \sum_{p \in \mathcal{P}}||\hat{I}_p(\Theta, \Phi, \mathcal{T})  - I_p||_1,
\end{align}
where $\mathcal{P}$ is the set of sampled pixels, which encourages the predicted intensity to match the intensity of the $p$th pixel in a training sonar image. The eikonal loss \cite{gropp2020implicit}
\begin{align}
    \mathcal{L}_{\text{eik}} \equiv \frac{1}{|\mathbf{X}|} \sum_{\spatialpoint\in \mathbf{X} } (||\nabla \mathbf{N}[\spatialpoint \left( \mathcal{T}) \right]||_2 - 1)^2,    
\end{align}
where $|\mathbf{X}|$ is the set of sampled 3D points, which is an implicit geometric regularization term used to regularize the SDF towards producing smooth reconstructions. Note that a 3D sample $\spatialpoint(\mathcal{T})$ is dependent on our current estimate of the pose parameters. Finally, we add an  $\ell_1$ loss term
\begin{align}
    \mathcal{L}_{\text{reg}} \equiv \frac{1}{|\mathbf{X}|} \sum_{\spatialpoint \in \mathbf{X}} || \alpha[\spatialpoint(\mathcal{T})]||_1,
\end{align}
to help produce favorable 3D reconstructions when we use sonar images from a limited set of view directions. Our final training loss term is therefore:
\begin{align}
\mathcal{L}(\Theta, \Phi, \mathcal{T}) = \mathcal{L}_{\text{int}} + \lambda_{\text{eik}} \mathcal{L}_{\text{eik}} + \lambda_{\text{reg}} \mathcal{L}_{\text{reg}}.
\label{lossfunction}
\end{align}

\noindent Our objective function is therefore: 
\begin{align}
    \Theta^*, \Phi^*, \mathcal{T}^* = \argmin_{\Theta, \Phi, \mathcal{T}} \mathcal{L}(\Theta, \Phi, \mathcal{T})
    \label{optimproblem}
\end{align}
Since our loss function is a fully differentiable function of all its parameters, we can perform parameter updates using iterative gradient-based methods such as ADAM. After convergence, we retrieve the surface by extracting the zero-level set of $\mathbf{N}$ using the Marching Cubes algorithm with a bounding box enclosing the object:
\begin{align}
    \mathcal{S} = \{\spatialpoint\in\mathbb{R}^3: \mathbf{N}(\spatialpoint) = 0 \}.
\end{align}

\subsection{Sensor Pose Parametrization}
\noindent Each drifting sonar pose $T_i$ is an element in $\text{SE}(3)$, the special Euclidean group in 3 dimensions. We parametrize it as a vector $(\mathbf{\omega}_i, \mathbf{t}_i) \in \text{se}(3)$ where $\mathbf{t}_i \in \mathbb{R}^3$ represents the translation and $\mathbf{\omega}_i \in \text{so}(3)$ represents the rotation in axis-angle form, and $\text{so}(3)$ is the Lie Algebra of rotations $\text{SO}(3)$. For each pose, we define a correction vector $(\delta\mathbf{\omega}_i, \delta\mathbf{t}_i) \in \mathbb{R}^6$ as a learnable parameter. After each gradient update, the full corrective matrix is then recovered via the exponential map:
\begin{align}
    \delta T_i = \begin{bmatrix}
    \delta \hat{\mathbf{\omega}} & \delta \mathbf{t} \\
    0 & 1
\end{bmatrix}
\end{align}
where $\delta \hat{\mathbf{\omega}}$ is the skew-symmetric matrix given by:
\begin{align}
   \delta \hat{\mathbf{\omega}} = \begin{bmatrix}
       0 & -\delta \omega_3 & \delta \omega_2 \\ 
       \delta \omega_3 & 0 & -\delta \omega_1 \\ 
       -\delta \omega_2 & \delta \omega_1 & 0
   \end{bmatrix} 
\end{align}
The final corrected pose transformation corresponding to image $i$ is then given by:
\begin{align}
    T_i \leftarrow T_i \cdot \delta T_i
\end{align}

\begin{table*}[h!]
% \vspace{2mm}
    \begin{center}
    \resizebox{0.6\paperwidth}{!}{
        \begin{tabular}{c|c|c|c|c|c|c|c|c|c|c}
            \toprule
            \multicolumn{2}{c|}{Datasets } & Real & Boat 1 & Boat 2 & Plane 1 & Plane 2 & Rock 1 & Rock 2 & Concrete column & Submarine\\
            \hline
            Elevation & $14\degree$ & 291 & 280 & 321 & 495 & 413 & 436 & 290 & 258 & 639 \\ 
            angle     & $28\degree$ & 441 & 283 & 492 & 444 & 364 & 435 & 289 & 258 & 639 \\
            \cline{1-2}
            \hline
    
        \end{tabular}
    }
    \end{center}
    \caption{Total number of poses in each dataset.}
    \label{table:num_frames}
    \vspace{-4mm}
\end{table*}

%%%%%%%%%%%%%%%%%%%%%%%%%%%%%%%%%%%%%%%%%%%%%%%%%%%%%%%%%%%%%%%%%%%%%%%%%%%%%%%%
\section{Evaluation}
\noindent  We trained our models on an NVIDIA H100 80GB HBM3 GPU with Intel Xeon 8470 CPU. Each training runs for $100$k iterations, which takes about 5 hours. Table \ref{table:num_frames} provides the total number of sonar/pose pairs for each simulated and real dataset. 

For our comparison metric, we use the mean and root mean square (RMS) Hausdorff distances. The Hausdorff distance is defined as:
\begin{align}
\begin{split}
    d_H(\mathcal{M}_1, \mathcal{M}_2) = \max ( \max_{\mathbf{p} \in \mathcal{M}_1} \min_{\mathbf{q} \in \mathcal{M}_2} || p - q ||_2 , \\ 
    \max_{\mathbf{q} \in \mathcal{M}_2} \min_{\mathbf{p} \in \mathcal{M}_1} || p - q ||_2  )
\end{split}
\end{align}
where $\mathcal{M}_1 $ and $\mathcal{M}_2$ are the ground-truth (GT) and reconstructed meshes respectively. 

\subsection{Understanding the Drift} 
\label{understandingthedrift}
\noindent  Similarly to ground robots, underwater vehicles often fuse measurements from multiple sensors to acquire accurate, drift-free odometry. Underwater vehicles are usually equipped with Doppler velocity logs (DVLs), inertial measurement units (IMUs), and depth sensors. A DVL provides low-noise measurements of vehicle velocity with respect to the sea floor in all three axes. An IMU typically consists of gyroscopes and accelerometers. By fusing DVL, IMU, and depth sensor measurements, an underwater vehicle is capable of providing noisy but drift-free measurements of Z-axis translation and pitch ($\phi$) and roll ($\theta$) angles by measuring hydrostatic pressure and direction of gravity \cite{Westman18tr}. On the other hand, the X and Y translations and yaw angle ($\psi$) are estimated from integration over gyroscope and DVL measurements, which will accumulate drift over time due to the absence of an absolute reference from the measurements.

 \begin{figure} [h!]
 \centering
 \includegraphics[width=1\columnwidth]{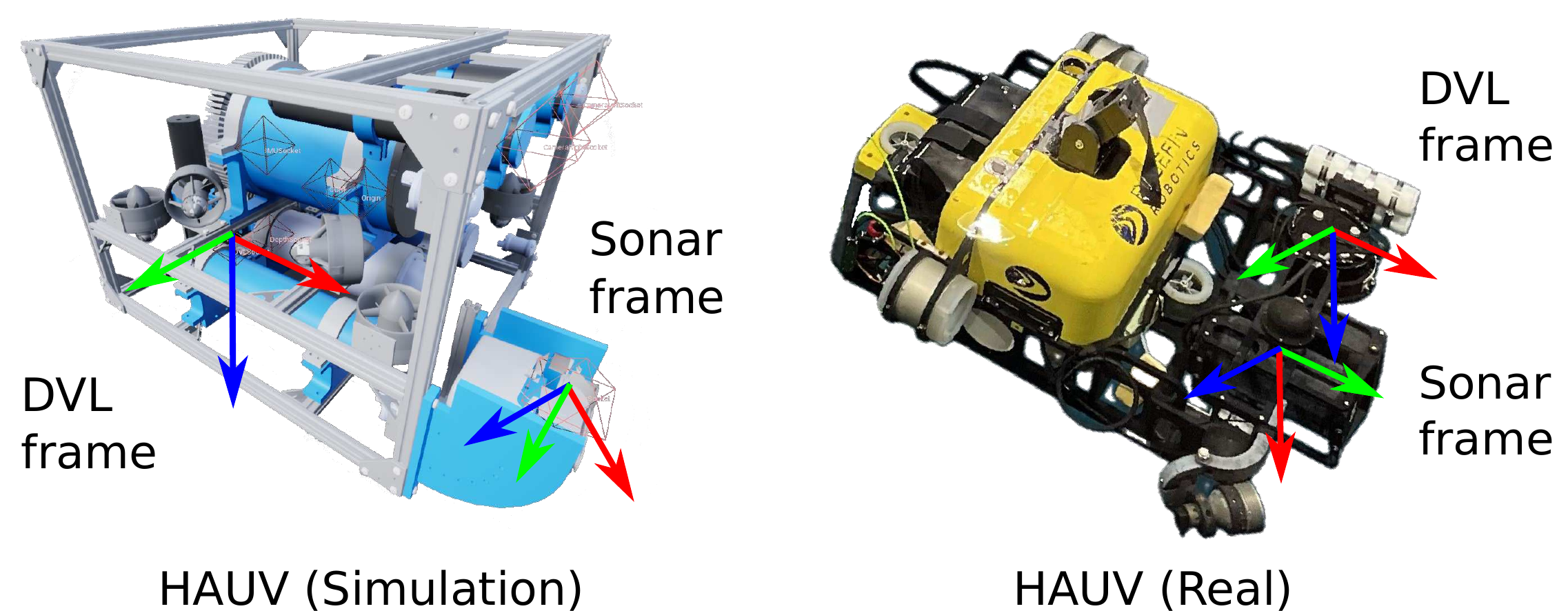}
\caption{Left: Simulated robot in HoloOcean \cite{HoloOceanDocs} with the DVL and an example sonar frame visualized. Right: Real HAUV with the DVL and sonar frames overlaid. Note that the DVL frame is similarly oriented in both the simulated and real setups.}
\label{fig:frame}
\vspace{-4mm}
\end{figure}

\subsection{Modeling the Drift}
\noindent Let $R_i=R_z(\psi_i)R_y(\phi_i)R_x(\theta_i)$ and $t_i=[x_i,y_i,z_i]^\top$ be the rotation matrix and translation vector of the 6DoF DVL pose $T_i$ at time $i$. We model the unbounded drift along the $x$, $y$, $\psi$ components of the AUV's odometric measurement as a stochastic process with a time-varying drift. Specifically, we assume that the difference between two consecutive DVL pose components, $u_i$ and $u_{i+1} \text{ for } u \in \{x, y, \psi\} $, follows:
\begin{align}
    &u_{i+1} - u_{i} \sim \mathcal{N}(\Delta u_{i,i+1}, q^u)
\end{align}
where $\Delta u_{i, i+1}$ represents the true underlying relative motion between timesteps $i$ and $i+1$ which depends on the AUV's control inputs. This can be rewritten as: 
\begin{align}
        &u_{i+1} - u_{i} = \Delta u_{i, i+1} + \varepsilon^u
        \label{noiseaddition}
\end{align}
where $\varepsilon^u \sim \mathcal{N}(0, q^u)$ is a zero-mean additive noise term with variance $q^u$ which models odometric uncertainty. A similar formulation for drifting poses was proposed by Westman et al. \cite{westman2019degeneracy}. The remaining components $u \in \{z, \theta, \phi \}$ are noisy but drift-free, and hence the noise in these components is modeled as zero-mean Gaussian noise. 

\subsection{Simulation}
\begin{table}[h!]
\vspace{2mm}
    \begin{center}
    \resizebox{1\linewidth}{!}{
        \begin{tabular}{c|c|c|c|?|c|c?c|c}
            %traj 6
            \toprule
            \multicolumn{2}{c|}{} &  \multicolumn{2}{c|}{NeuSIS (GT)}&  \multicolumn{2}{c|}{NeuSIS (drift)}&  \multicolumn{2}{c}{Ours}  \\
            \cline{3-8}
            \multicolumn{2}{c|}{} &  RMS & Mean &  RMS & Mean &  RMS & Mean  \\
            \hline
    Boat 1                        & $14\degree$  & 0.074 & 0.059 & 0.101 & 0.076 & \textbf{0.089} & \textbf{0.065}\\
    $(3.8 \times 1.7\times 0.84)$ & $28\degree$  & 0.065 & 0.049 & 0.102 & 0.076 & \textbf{0.067} & \textbf{0.049}\\
         \hline
    Boat 2                        & $14\degree$  & 0.093 & 0.064 & 0.146 & 0.100 & \textbf{0.098} & \textbf{0.070}\\
    $(5.7 \times 2.3\times 1.2)$  & $28\degree$  & 0.108 & 0.079 & 0.179 & 0.127 & \textbf{0.109} & \textbf{0.082}\\
         \hline
    Plane 1                       & $14\degree$  & 0.156 & 0.102 & 0.197 & 0.141 & \textbf{0.159} & \textbf{0.122}\\
    $(13.5 \times 11.5\times 3.6)$& $28\degree$  & 0.175 & 0.121 & 0.217 & 0.153 & \textbf{0.183} & \textbf{0.126}\\
         \hline
    Plane 2                       & $14\degree$  & 0.117 & 0.091 & 0.400 & 0.162 & \textbf{0.151} & \textbf{0.115}\\
    $(9.1 \times 12.6\times 3.0)$ & $28\degree$  & 0.150 & 0.115 & 1.551 & 0.737 & \textbf{0.266} & \textbf{0.194}\\
         \hline
    Rock 1                        & $14\degree$  & 0.139 & 0.105 & 0.194 & 0.150 & \textbf{0.154} & \textbf{0.113}\\
    $(5.7 \times 3.5 \times 2.8)$ & $28\degree$  & 0.112 & 0.082 & 0.196 & 0.148 & \textbf{0.129} & \textbf{0.098}\\    
         \hline
    Rock 2                        & $14\degree$  & 0.110 & 0.083 & 0.148 & 0.112 & \textbf{0.117} & \textbf{0.091}\\
    $(2.2 \times 2.2 \times 2.0)$ & $28\degree$  & 0.135 & 0.102 & 0.169 & 0.127 & \textbf{0.151} & \textbf{0.113}\\
         \hline
    Concrete column               & $14\degree$  & 0.058 & 0.033 & 0.108 & 0.074 & \textbf{0.080} & \textbf{0.059}\\
    $(1.9 \times 1.2 \times 4.3)$ & $28\degree$  & 0.055 & 0.038 & 0.784 & 0.057 & \textbf{0.052} & \textbf{0.039}\\
         \hline
    Submarine                     & $14\degree$  & 0.149 & 0.116 & 0.206 & 0.155 & \textbf{0.150} & \textbf{0.117}\\
    $(5.1\times 16.7 \times 4.7)$ & $28\degree$  & 0.144 & 0.110 & 0.280 & 0.209 & \textbf{0.186} & \textbf{0.141}\\
         \hline
        \end{tabular}
    }
    \end{center}
    \caption{Size (\(W \times L \times H\)), root mean square (RMS), and mean Hausdorff distance (meters) for eight different objects from HoloOcean. The results show that our method produces more accurate 3D reconstructions compared to NeuSIS with drifting poses, demonstrating the effectiveness of our approach in handling pose drift.}
    \label{table:nerfmm_simtable}
    \vspace{-2mm}
\end{table}

\noindent  We used an imaging sonar dataset of objects of different shapes and sizes collected using HoloOcean \cite{Potokar24joe}, an underwater simulator. The dataset was collected with the simulation of multipath effects enabled and the inclusion of multiplicative noise $w^{\text{sm}} \sim \mathcal{N}(0,0.15)$ and additive noise $w^{\text{sa}} \sim \mathcal{R}(0.2)$ where $\mathcal{R}$ is the Rayleigh distribution. The original dataset was collected with a frequency of 10Hz, which we downsample by a factor of two.
Among other sensors, the simulated robot is equipped with a DVL whose frame is oriented similarly to the DVL on the real robot in Fig.~\ref{fig:frame}. For each object, we collect two to three different trajectories while varying sonar orientation. For each sonar image, HoloOcean additionally returns the ground-truth sonar and DVL poses. Hence, to simulate realistic drifting pose patterns as described in subsection \ref{understandingthedrift}, we adopt the following strategy: Let $T_i^\text{dvl}$ and $T_i^\text{sonar}$ be the ground-truth DVL and sonar poses at time $i$ respectively. 
\begin{enumerate}
    \item Compute the relative DVL pose between timesteps $i$ and $i+1$: $\Delta T_{i\rightarrow i+1}^\text{dvl} = (T_{i}^\text{dvl})^{-1} \cdot T_{i+1}^\text{dvl}$.
    \item Add noise to the $x, y, \psi$ axes of $\Delta T_{i, i+1}^\text{dvl}$ following Eq.~\ref{noiseaddition} with $\varepsilon^x, \varepsilon^y \sim \mathcal{N}(0, 0.004$m$)$ and $\varepsilon^\psi \sim \mathcal{N}(0, 0.004\text{rad})$. We obtain the noisy relative transform $\Delta \tilde{T}_{i\rightarrow i+1}^\text{dvl}$.
    \item Compute the noisy DVL pose at timestep $i+1$ as $\tilde{T}_{i+1}^\text{dvl} = T_{i}^\text{dvl} \cdot \Delta \tilde{T}_{i\rightarrow i+1}^\text{dvl}$
    \item To simulate the noisy but drift-free measurements over the $z, \phi, \theta$ axes, we add Gaussian noise to $\tilde{T}_{i+1}^\text{dvl}$ to each of these axes with $\varepsilon^z \sim \mathcal{N}(0,0.005\text{m})$ and $\varepsilon^\phi, \varepsilon^\theta \sim \mathcal{N}(0,0.005\text{rad})$.
    \item Obtain the corresponding noisy sonar pose by multiplying with the known DVL-to-sonar extrinsic matrix: $\tilde{T}_{i+1}^\text{sonar} = \tilde{T}_{i+1}^\text{dvl} \cdot T_{\text{dvl}\rightarrow\text{sonar}} $.
\end{enumerate}

\begin{figure*}[h!]
    \centering
    \begin{subfigure}[b]{0.48\textwidth}
        \centering
        \includegraphics[width=\linewidth]{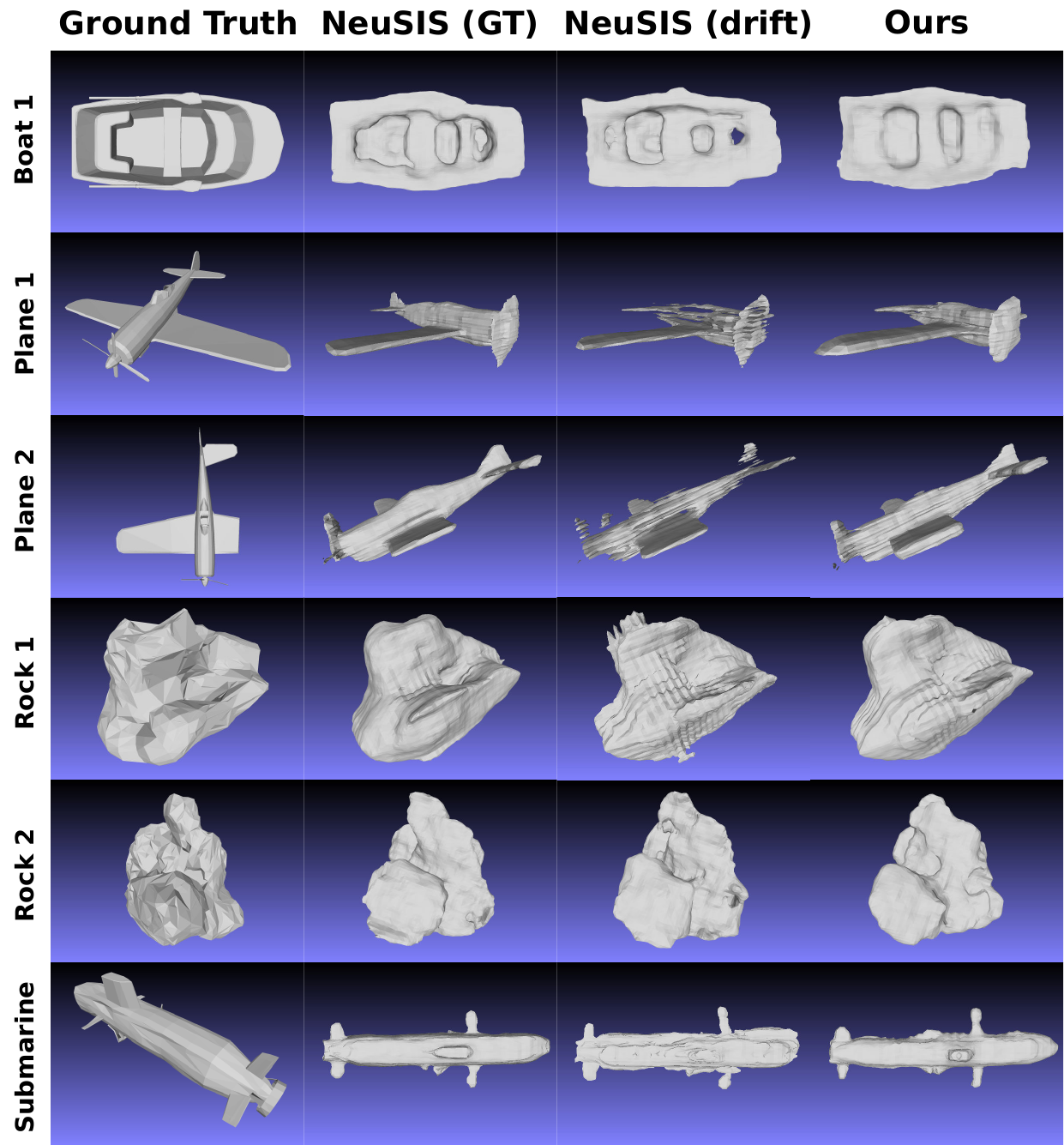}
        \caption{3D reconstruction results for the \(14^\circ\) elevation aperture simulated sonar datasets.}
        \label{fig:sim14}
    \end{subfigure} \hfill
    \begin{subfigure}[b]{0.48\textwidth}
        \centering
        \includegraphics[width=\linewidth]{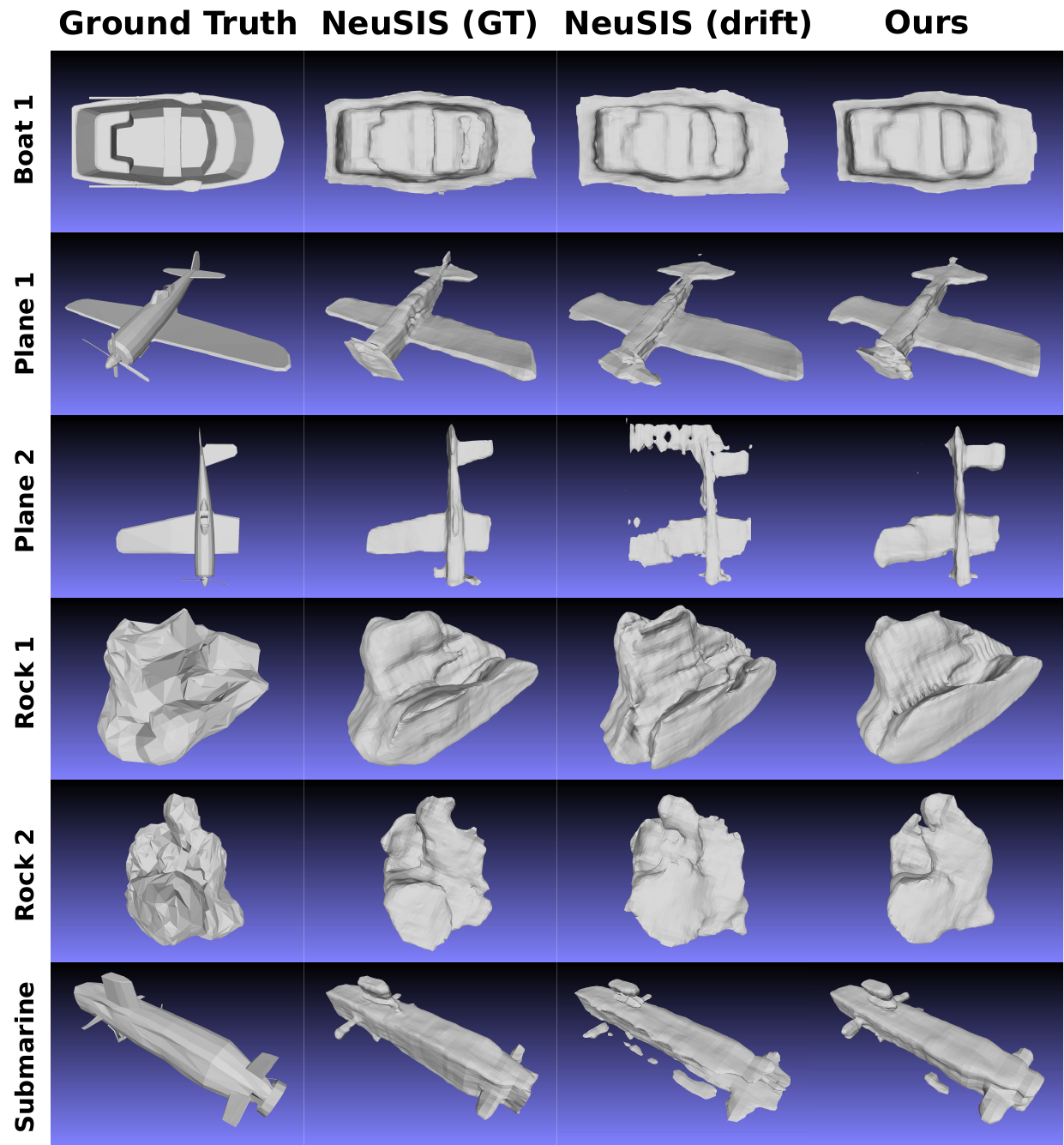}
        \caption{3D reconstruction results for the \(28^\circ\) elevation aperture simulated sonar datasets.}
        \label{fig:sim28}
    \end{subfigure}
    \caption{3D reconstruction results for (a) the \(14^\circ\) and (b) the \(28^\circ\) elevation aperture sonar datasets collected using the HoloOcean underwater simulator. From left to right, the images show ground-truth meshes of six different objects, followed by reconstructions from NeuSIS with ground-truth odometry, NeuSIS with drifting poses, and our proposed method. Our approach effectively restores dense 3D reconstructions despite drifting odometry, achieving results comparable to NeuSIS with ground-truth trajectories.}
    \label{fig:sim_results}
    \vspace{-4mm}
\end{figure*}

Figs.~\ref{fig:sim14} and \ref{fig:sim28} present qualitative results for different simulated objects at elevation apertures of 14$^\circ$ and 28$^\circ$. We compare (1) reconstructions using ground-truth (GT) odometry poses: NeuSIS (GT), (2) reconstructions with noisy pose measurements without optimization: NeuSIS (drift), and (3) our method, which jointly optimizes the SDF, renderer, and poses. For each object and each method, we select the best level set, i.e. Marching Cubes threshold, $\epsilon  \in [-0.2, 0.2]$.

As expected, the best results are achieved using the GT sonar poses from the HoloOcean simulator. These reconstructions represent an upper bound on reconstruction quality for each object. Our goal is to approach the accuracy of NeuSIS (GT), both qualitatively and quantitatively, even after injecting noise in the pose measurements. Qualitatively, our method consistently reduces reconstruction errors compared to using drifting poses across all objects. Notably, we observe less stratification in the \textit{Plane 1} and \textit{Plane 2} datasets at 14$^\circ$, and a significant correction in the \textit{Plane 2} dataset at 28$^\circ$—particularly in the tail area. Similarly, the \textit{submarine  }reconstructions at 14$^\circ$ and 28$^\circ$ exhibit improvements along the entire shape. 
These qualitative results are supported by the quantitative results in Table \ref{table:nerfmm_simtable} which demonstrate a significant improvement in reconstruction accuracy when using our method.

\subsection{Water Tank Experiments}
\begin{figure}[h!]
    \begin{subfigure}[h]{0.24\textwidth}
        \centering
        \includegraphics[width=0.8\columnwidth]{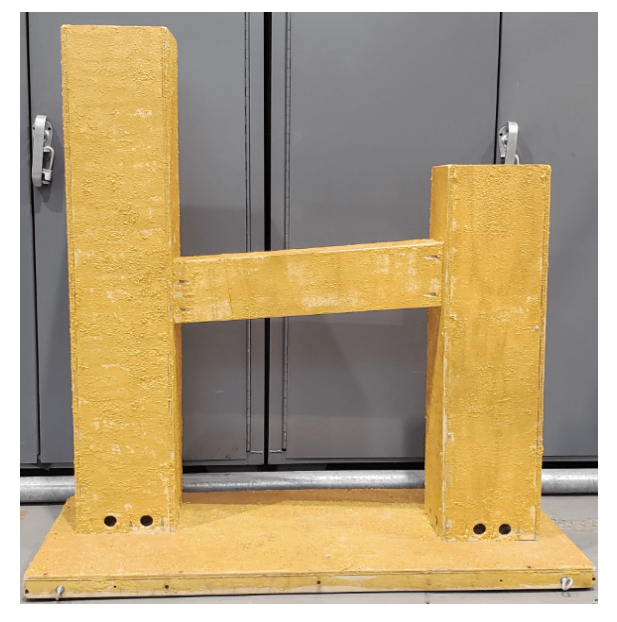}
        \caption{Test structure.}
        \label{realstructure}
    \end{subfigure}
    \begin{subfigure}[h]{0.24\textwidth}
        \centering
        \includegraphics[width=0.8\columnwidth]{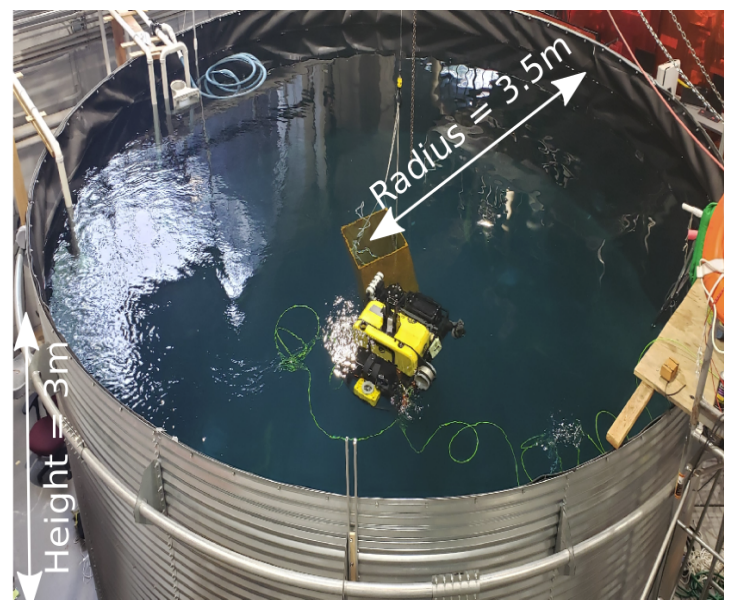}
        \label{testtank}
        \caption{Water test tank.}
    \end{subfigure}%
    \caption{Real-world experimental setup.}
    \label{fig:Realsetup}
    \vspace{-4mm}
\end{figure}
\noindent We evaluate our proposed method on real-world datasets of a test structure submerged in a test tank 
(see Fig.~\ref{fig:Realsetup}) imaged with the two wide elevation apertures achievable by the sonar: $14^\circ$ and $28^\circ$. 
Our experimental platform for dataset collections is a Bluefin Hovering Autonomous Underwater Vehicle (HAUV) \cite{BluefinHAUV} equipped with a 1.2MHz Teledyne/RDI Workhorse Navigator Doppler velocity log (DVL), a Honeywell HG1700 inertial measurement unit (IMU), a Paroscientific Digiquartz depth sensor, and a Sound Metric DIDSON imaging sonar \cite{DIDSON} (please check \cite{westman2020theory, qadri23icra} for more details about the datasets and the collection setup). The frame rate of the real-world datasets is 2Hz.

\begin{figure} [b!]
    \centering
    \includegraphics[width=1.05\columnwidth]{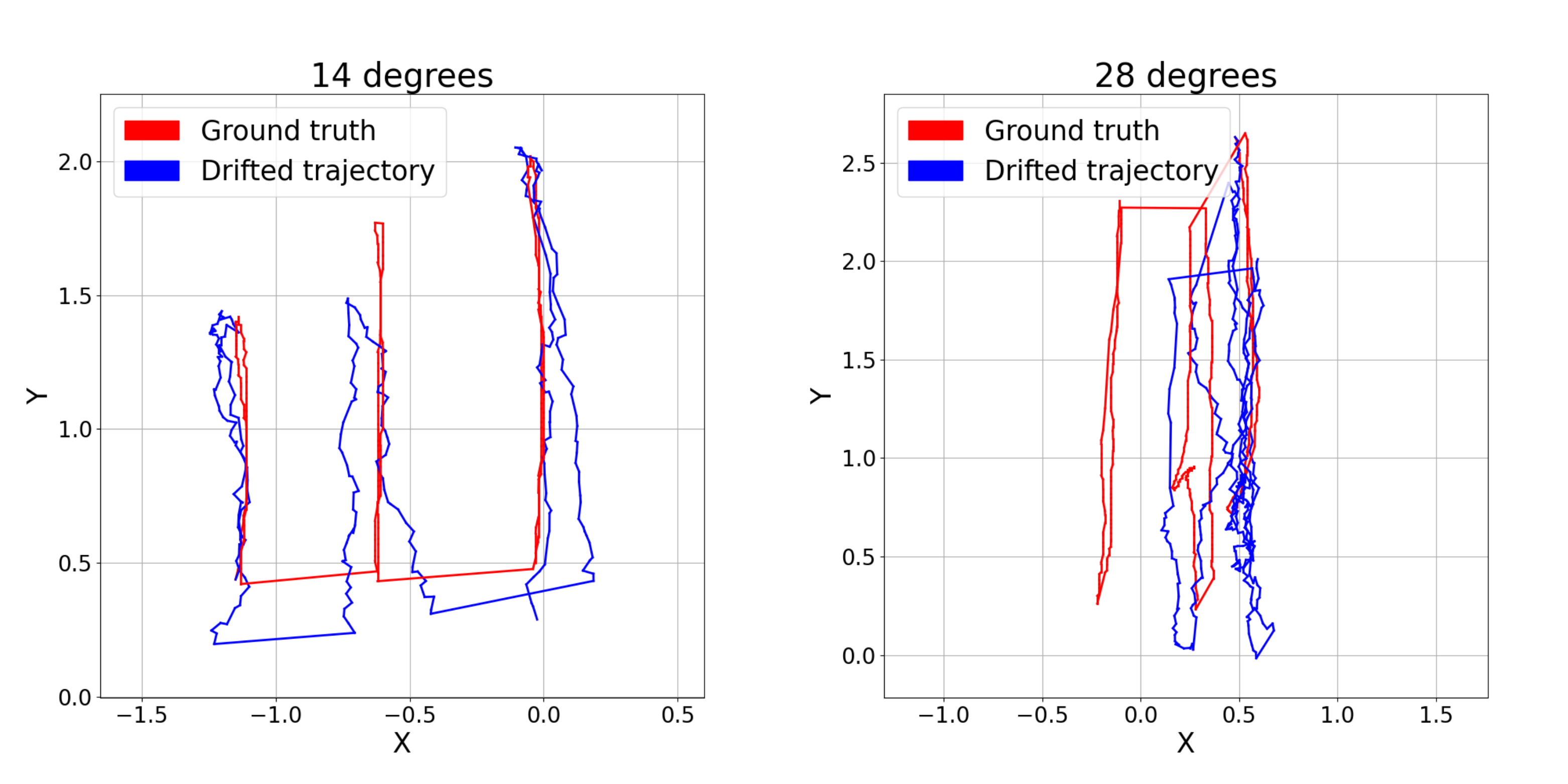}
    \caption{Top-down view illustrating an example of a drifting DVL trajectory after noise injection (blue) alongside its corresponding DVL poses with no added noise (red) for the \(14^\circ\) and \(28^\circ\) elevation aperture real datasets, shown on the left and right, respectively. Noise is injected into the \(x\), \(y\), and \(\psi\) relative poses with \(\varepsilon^x, \varepsilon^y \sim \mathcal{N}(0, 0.015\ \text{m})\) and \(\varepsilon^\psi \sim \mathcal{N}(0, 0.015\ \text{rad})\).}
    \label{fig:sampletraj}
    \vspace{-2mm}
\end{figure}

\begin{figure*}[h!]
    \centering
    \begin{subfigure}[b]{0.48\textwidth}
        \centering
        \includegraphics[width=\linewidth]{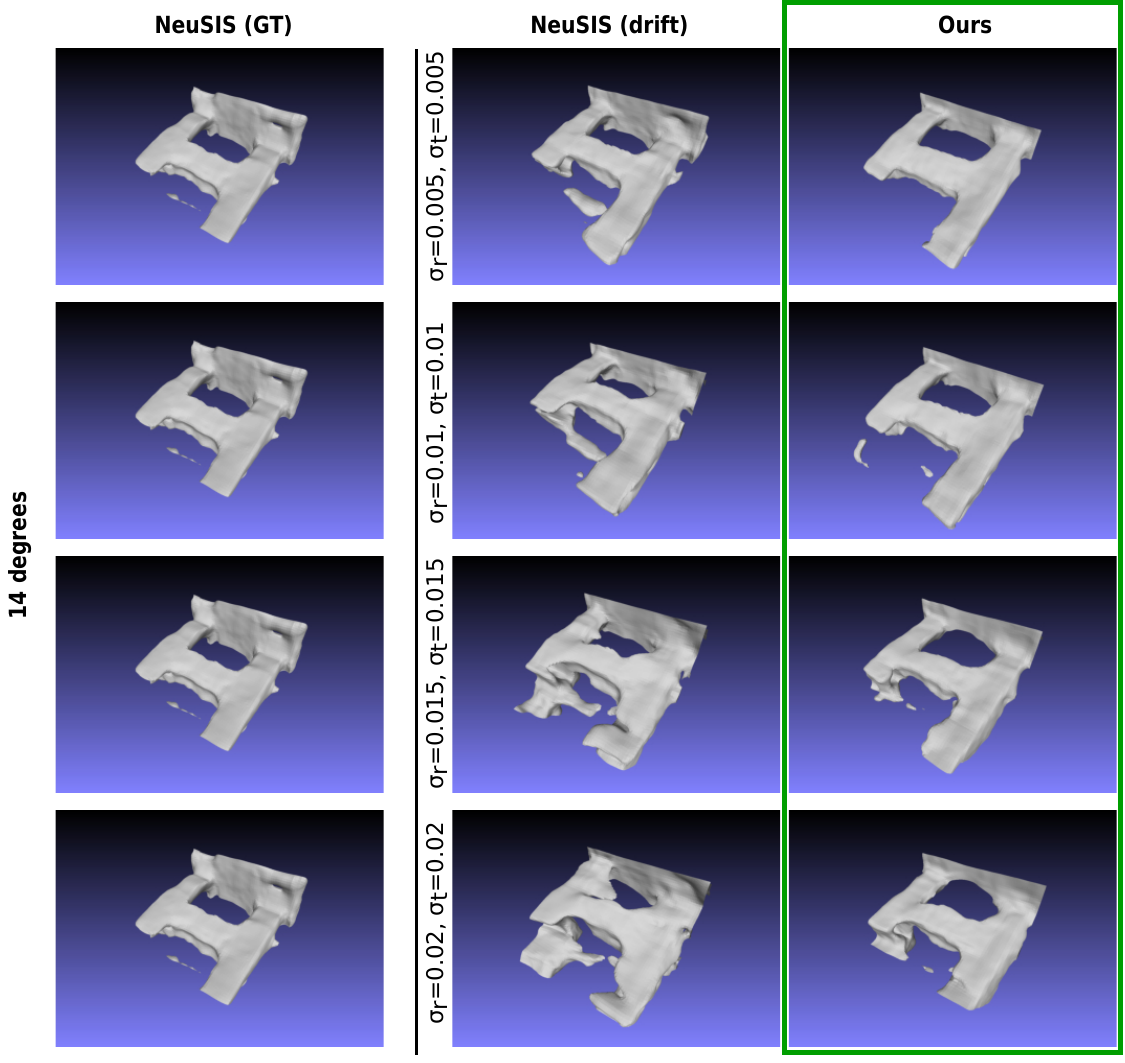}
        \caption{3D reconstruction results for the \(14^\circ\) elevation aperture real sonar datasets.}
        \label{fig:real14}
    \end{subfigure} \hfill
    \begin{subfigure}[b]{0.48\textwidth}
        \centering
        \includegraphics[width=\linewidth]{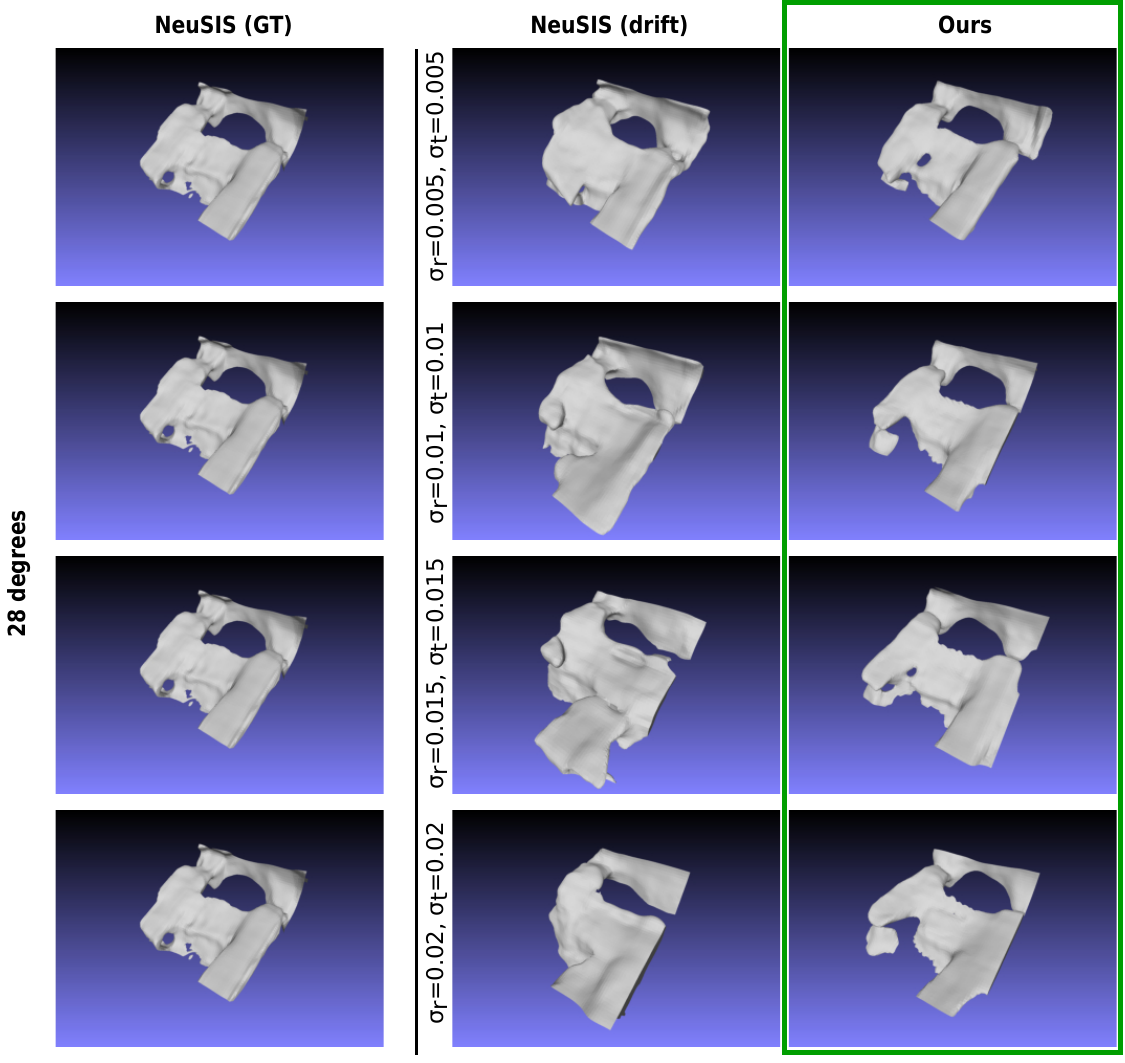}
        \caption{3D reconstruction results for the \(28^\circ\) elevation aperture real sonar datasets.}
        \label{fig:real28}
    \end{subfigure}
    \caption{3D reconstructions from each method with two different elevation apertures and four sets of drifting noise added to the \(x\), \(y\), and yaw directions of the vehicle odometry. Our proposed method yields more accurate and cleaner reconstructions compared to NeuSIS with drifting noise added to the vehicle odometry. Notably, reconstructions from the proposed method with lower drifting noise are cleaner than those from NeuSIS with ground-truth odometry, demonstrating the capability of the proposed method to eliminate drifting noise in the odometry.}
    \label{fig:realQualitativeNeRFmm}
    \vspace{-4mm}
\end{figure*}

 We start by discarding sonar image/pose pairs that lack returns from the object of interest. We then inject relative pose noise as described in subsection \ref{noiseaddition}, adding $\varepsilon^x, \varepsilon^y \sim \mathcal{N}(0, \sigma_t)$ to the relative DVL $x$ and $y$ measurements and $\varepsilon^\psi \sim \mathcal{N}(0, \sigma_r)$ to the relative yaw ($\psi$). Our method is evaluated across four increasing noise levels (listed in Table \ref{table:nerfmm_realtable}), with each experiment (i.e., each elevation angle and noise level) repeated using three different random seeds to simulate different noise patterns. 
Fig.~\ref{fig:sampletraj} shows an example of a DVL trajectory before and after noise injection.

\begin{table}[h!]
    \begin{center}
    \resizebox{0.6\linewidth}{!}{
        \begin{tabular}{c|c?c|c}
            \toprule
            \multicolumn{1}{c|}{} &  \multicolumn{3}{c}{NeuSIS (GT)}  \\
            \cline{2-4}
            \multicolumn{1}{c|}{} & elevation &  RMS & Mean  \\
            \hline
    Real   & $14\degree$ & \textbf{0.052} & \textbf{0.036} \\
    Datasets & $28\degree$ & \textbf{0.071} & \textbf{0.049} \\
        \hline
        \end{tabular}
    }
    \end{center}
    \caption{Root mean square (RMS) and mean Hausdorff distance in meters for NeuSIS with ground-truth odometry on real-world datasets with \(14^\circ\) and \(28^\circ\) elevation apertures.}
    \label{table:realtable}
    \vspace{-4mm}
\end{table}

\begin{table}[h!]
    \begin{center}
    \resizebox{1\linewidth}{!}{
        \begin{tabular}{c|c|c|c?c|c}
            \toprule
            \multicolumn{2}{c|}{} &  \multicolumn{2}{c|}{NeuSIS (drift)}&  \multicolumn{2}{c}{Ours}  \\
            \cline{3-6}
            \multicolumn{2}{c|}{} & RMS & Mean &  RMS & Mean   \\
            \hline
    $\sigma_r=0.005\,\text{m}$   & $14\degree$ & 0.052 $\pm$ 0.001 & 0.039 $\pm$ 0.002 & \textbf{0.046 $\pm$ 0.001} & \textbf{0.034 $\pm$ 0.001} \\
    $\sigma_t=0.005\,\text{rad}$ & $28\degree$ & 0.082 $\pm$ 0.003 & 0.058 $\pm$ 0.002 & \textbf{0.073 $\pm$ 0.002} & \textbf{0.052 $\pm$ 0.001} \\
        \hline
    $\sigma_r=0.01\,\text{m}$    & $14\degree$ & 0.061 $\pm$ 0.001 & 0.046 $\pm$ 0.004 & \textbf{0.047 $\pm$ 0.005} & \textbf{0.035 $\pm$ 0.003} \\
    $\sigma_t=0.01\,\text{rad}$  & $28\degree$ & 0.085 $\pm$ 0.002 & 0.062 $\pm$ 0.000 & \textbf{0.074 $\pm$ 0.002} & \textbf{0.054 $\pm$ 0.001} \\
    \hline
    $\sigma_r=0.015\,\text{m}$   & $14\degree$ & 0.073 $\pm$ 0.001 & 0.056 $\pm$ 0.003 & \textbf{0.054 $\pm$ 0.003} & \textbf{0.040 $\pm$ 0.003} \\
    $\sigma_t=0.015\,\text{rad}$ & $28\degree$ & 0.091 $\pm$ 0.002 & 0.069 $\pm$ 0.002 & \textbf{0.079 $\pm$ 0.004} & \textbf{0.056 $\pm$ 0.002} \\
        \hline
    $\sigma_r=0.02\,\text{m}$    & $14\degree$ & 0.081 $\pm$ 0.005 & 0.062 $\pm$ 0.006 & \textbf{0.062 $\pm$ 0.004} & \textbf{0.047 $\pm$ 0.004} \\
    $\sigma_t=0.02\,\text{rad}$  & $28\degree$ & 0.093 $\pm$ 0.004 & 0.069 $\pm$ 0.004 & \textbf{0.081 $\pm$ 0.003} & \textbf{0.060 $\pm$ 0.004} \\        
        \hline
        \end{tabular}
    }
    \end{center}
    \caption{Root mean square (RMS) and mean Hausdorff distances for reconstructions of \(14^\circ\) and \(28^\circ\) real-world datasets using NeuSIS with drifting odometry and the proposed method. Translation is measured in meters and rotation in radians. We select three random seeds and compute the mean and standard deviation for each distance. The Hausdorff distance threshold is set to \(0.2\) m for the \(14^\circ\) elevation dataset and \(0.25\) m for the \(28^\circ\) dataset.}
    \label{table:nerfmm_realtable}
    \vspace{-3mm}
\end{table}
Fig.~\ref{fig:realQualitativeNeRFmm} shows qualitative results of reconstructions using the 4 different noise level and the 2 elevation apertures.
We observe that for both elevation apertures, the reconstructions with NeuSIS (drift) degrade rapidly as the noise level increases. For example, at $14^\circ$, increasing artifacts can be observed near the shorter leg region, while at $28^\circ$, we observe the deterioration of the entire shape. In contrast, our method successfully recovers shapes in which the main components of the structure
are preserved (a base, a smaller and larger leg,
and a crossbar). 
We report in Table \ref{table:realtable}, the result obtained using NeuSIS (GT) -- i.e. reconstructions using the GT odometry. Table \ref{table:nerfmm_realtable} shows the metrics (average $\pm$ standard deviation) for both our method as well as NeuSIS (drift) after noise injection. As expected, the reconstruction quality of NeuSIS (drift) degrades with increasing noise. However, our method remains robust to significant pose drift, and only begins to struggle when the noise becomes particularly high ($\sigma_r = 0.02\text{rad}$ and $\sigma_t = 0.02\text{m}$) between consecutive relative poses.

\section{Do we Recover the Odometry Poses?}

\noindent We experimentally demonstrate that the set of possible poses, $\mathcal{T}$, that minimizes the reconstruction error is not unique. In other words, our algorithm can converge to a set of poses that is far from the odometry poses while still resulting in final 3D reconstructions that are perceptibly similar. We perform the following experiment on the $28^\circ$ real data:
\begin{enumerate}
    \item \textbf{Train using odometry poses:} Use the odometry poses, $\mathcal{T}_{\text{odom}}$,  and train a 3D model $\mathcal{R}_{\text{odom}}$ until converge. 
    \item \textbf{Freeze network weights:} Freeze the weights of both the SDF network, $\textbf{N}$, and the neural renderer, $\textbf{M}$.
    \item \textbf{Inject noise to relative pose: } Corrupt the odometry poses by injecting noise to the $x, y, \psi$ relative poses with $\varepsilon^x, \varepsilon^y \sim \mathcal{N}(0, 0.01$m$)$ and $\varepsilon^\psi \sim \mathcal{N}(0, 0.01\text{rad})$. We obtain a noisy drifting set of poses: $\mathcal{T}_{\text{noisy}}$.
    \item \textbf{Optimize the noisy poses:} Optimize $\mathcal{T}_{\text{noisy}}$ using the loss function in Eq.~\ref{lossfunction} while keeping the weights of networks $\textbf{N}$ and $\textbf{M}$ from step 2 frozen. In other words, solve the following optimization problem:
    \begin{align*}
    \mathcal{T}^*_{\text{noisy}} = \argmin_{\mathcal{T}} \mathcal{L}(\Theta, \Phi, \mathcal{T})
\end{align*}
\item \textbf{Train a new 3D model:} Finally, freeze $\mathcal{T}^*_{\text{noisy}}$ and train a new 3D model $\mathcal{R}_{\text{noisy}}^*$.
\end{enumerate}

\begin{figure} [h!]
 \centering
 \vspace{-3mm}
 \includegraphics[width=1\columnwidth]{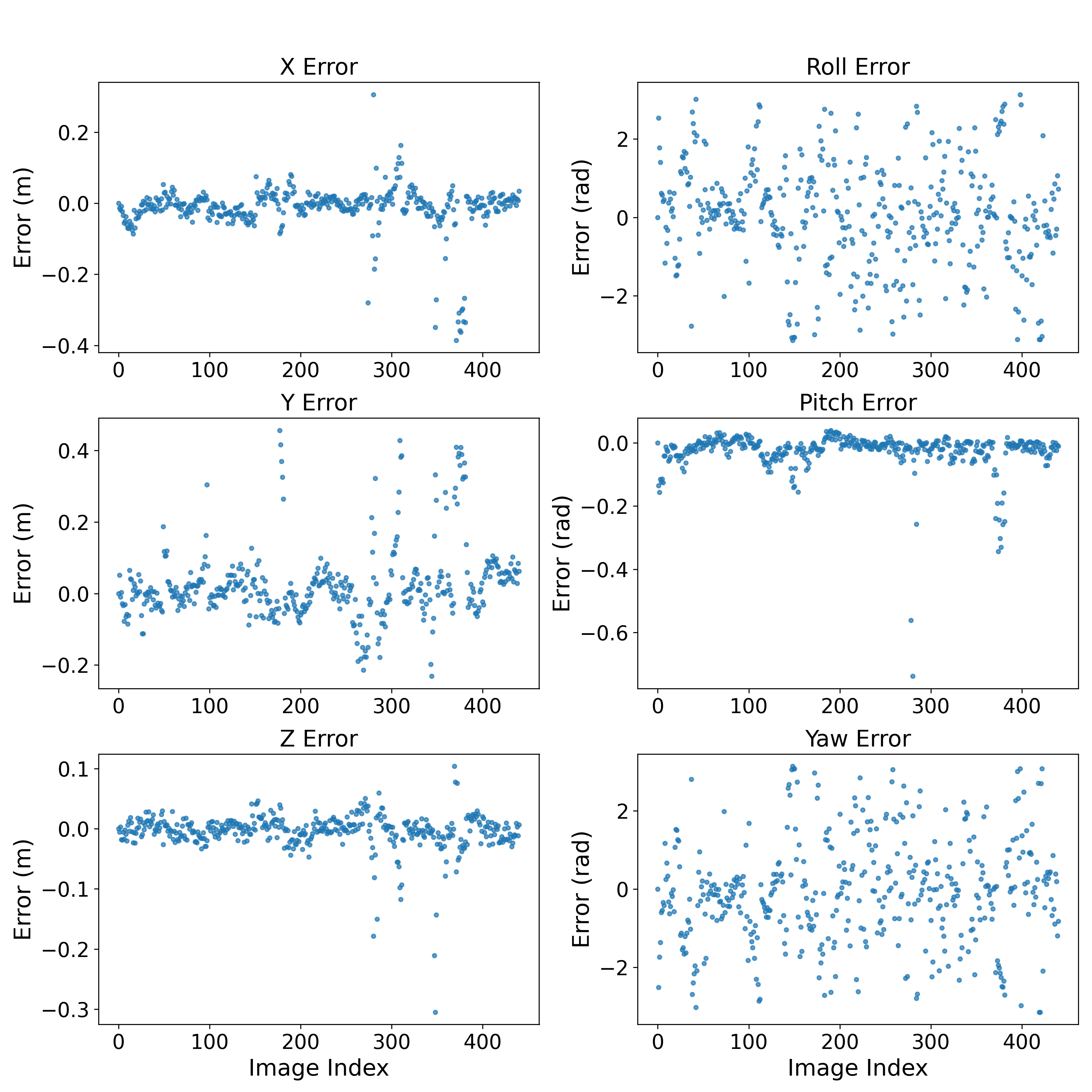}
\caption{Error in each pose component between the odometry poses, $\mathcal{T}_\text{odom}$ and the optimized poses after noise injection, $\mathcal{T}_\text{noisy}^*$. We note that the two sets of poses differ significantly. }
\label{fig:poseerrors}
\vspace{-4mm}
\end{figure}
\begin{wrapfigure}{r}{0.34\columnwidth}
  \vspace{-1.8em}
  \begin{center}
    \includegraphics[width=0.32\columnwidth]{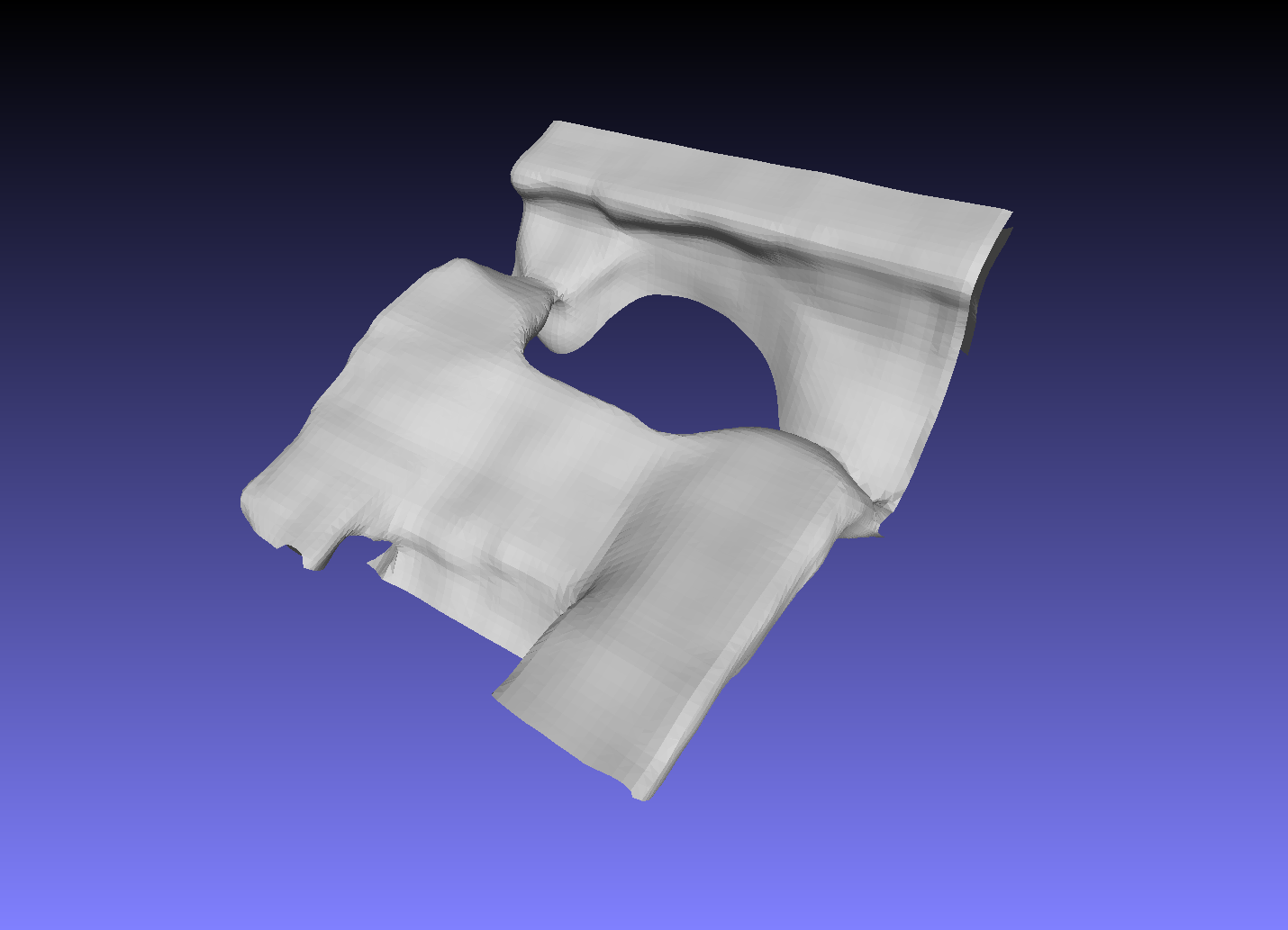}
  \end{center}
%  \vspace{-1.2em}
  \caption{The reconstruction $\mathcal{R}_\text{noisy}^*$ obtained from step  5: Freeze $\mathcal{T}_{\text{noisy}}^*$ and train $\mathbf{M}$ and $\mathbf{N}$. }
  \label{fig:sectionivoptimized}
\end{wrapfigure}
Fig.~\ref{fig:poseerrors} shows the errors between $\mathcal{T}_{\text{odom}}$ and $\mathcal{T}_{\text{noisy}}^*$ in the sonar frame, revealing significant differences between the two sets of poses. 
This indicates that if we freeze the converged mesh $\mathcal{R}_{\text{odom}}$ and optimize only the poses, the optimization can still converge to a solution set far from the odometric poses.
Fig.~\ref{fig:sectionivoptimized} shows the resulting mesh reconstruction when we freeze $\mathcal{T}_\text{noisy}^*$ and train $\mathbf{M}$ and $\mathbf{N}$ to obtain $\mathcal{R}_\text{noisy}^*$. The reconstruction $\mathcal{R}^*_{\text{noisy}}$ captures the major parts of the structure (two legs and the middle tile), supporting the observation that even significantly different pose sets can produce plausible 3D reconstructions. This result is supported by the quantitative metrics for $\mathcal{R}_\text{noisy}^*$ (RMS = $0.077$m, mean = $0.058$m), which are close to the metrics reported when using $\mathcal{T}_\text{odom}$ in Table \ref{table:realtable}.

%%%%%%%%%%%%%%%%%%%%%%%%%%%%%%%%%%%%%%%%%%%%%%%%%%%%%%%%%%%%%%%%%%%%%%%%%%%%%%%%
\section{Conclusion and Future Work}
\noindent 
We proposed an approach for reconstructing objects using imaging sonar, even in the presence of significant pose drift. Our method jointly optimizes both the sonar poses and the 3D model parameters using only the reconstruction loss, without relying on external sensors. That is, it learns directly using only the sonar images and their corresponding noisy poses. Through extensive experiments across different objects and elevation apertures, we demonstrated that our method is robust to high noise levels and adapts well to diverse target geometries.

For future work, we see multiple promising directions. First, incorporating additional sensor modalities available on the vehicle, such as Doppler Velocity Logs (DVL), IMUs, or optical cameras, could provide stronger constraints for pose optimization, further improving reconstruction accuracy. Second, our current approach is designed for offline 3D reconstruction. To enable real-time applications, we plan to explore acceleration techniques such as Instant-NGP \cite{instantngp}, which uses neural graphics primitives for highly efficient scene optimization. These enhancements would make our method more practical for real-world autonomous underwater operations.

%%%%%%%%%%%%%%%%%%%%%%%%%%%%%%%%%%%%%%%%%%%%%%%%%%%%%%%%%%%%%%%%%%%%%%%%%%%%%%%%
\section{Acknowledgment}
\noindent
T.L., M.Q.,~and M.K. were partially supported by the Office of Naval Research (ONR) grant N00014-24-1-2272.  K.Z.~and C.A.M.~were partially supported by AFOSR award no.~FA9550-22-1-0208, NSF award no.~2339616, and ONR award no.~N000142312752.

%%%%%%%%%%%%%%%%%%%%%%%%%%%%%%%%%%%%%%%%%%%%%%%%%%%%%%%%%%%%%%%%%%%%%%%%%%%%%%%%
% \section*{Appendix}
\bibliographystyle{IEEEtran}
\IEEEtriggeratref{47}
\bibliography{main}

\begin{thebibliography}{10}
\providecommand{\url}[1]{#1}
\csname url@samestyle\endcsname
\providecommand{\newblock}{\relax}
\providecommand{\bibinfo}[2]{#2}
\providecommand{\BIBentrySTDinterwordspacing}{\spaceskip=0pt\relax}
\providecommand{\BIBentryALTinterwordstretchfactor}{4}
\providecommand{\BIBentryALTinterwordspacing}{\spaceskip=\fontdimen2\font plus
\BIBentryALTinterwordstretchfactor\fontdimen3\font minus \fontdimen4\font\relax}
\providecommand{\BIBforeignlanguage}[2]{{%
\expandafter\ifx\csname l@#1\endcsname\relax
\typeout{** WARNING: IEEEtran.bst: No hyphenation pattern has been}%
\typeout{** loaded for the language `#1'. Using the pattern for}%
\typeout{** the default language instead.}%
\else
\language=\csname l@#1\endcsname
\fi
#2}}
\providecommand{\BIBdecl}{\relax}
\BIBdecl

\bibitem{wang2019underwater}
J.~Wang, T.~Shan, and B.~Englot, ``Underwater terrain reconstruction from forward-looking sonar imagery,'' in \emph{Proc. IEEE Intl. Conf. on Robotics and Automation (ICRA)}, Montreal, Canada, May 2019, pp. 3471--3477.

\bibitem{negahdaripour2018application}
S.~Negahdaripour, ``Application of forward-scan sonar stereo for 3-{D} scene reconstruction,'' \emph{IEEE J. of Oceanic Engineering (JOE)}, vol.~45, no.~2, pp. 547--562, Oct. 2018.

\bibitem{albiez2015flatfish}
J.~Albiez, S.~Joyeux, C.~Gaudig, J.~Hilljegerdes, S.~Kroffke, C.~Schoo, S.~Arnold, G.~Mimoso, P.~Alcantara, R.~Saback \emph{et~al.}, ``Flatfish-a compact subsea-resident inspection {AUV},'' in \emph{Proc. IEEE/MTS OCEANS Conf. and Exhibition}, DC, USA, Oct. 2015, pp. 1--8.

\bibitem{Lin23icra}
T.~Lin, A.~Hinduja, M.~Qadri, and M.~Kaess, ``Conditional {GAN}s for sonar image filtering with applications to underwater occupancy mapping,'' in \emph{Proc. IEEE Intl. Conf. on Robotics and Automation (ICRA)}, London, UK, May 2023, pp. 1048--1054.

\bibitem{qadri23icra}
M.~Qadri, M.~Kaess, and I.~Gkioulekas, ``Neural implicit surface reconstruction using imaging sonar,'' in \emph{Proc. IEEE Intl. Conf. on Robotics and Automation (ICRA)}, London, UK, May 2023, pp. 1040--1047.

\bibitem{aykin2016three}
M.~D. Aykin and S.~Negahdaripour, ``Three-dimensional target reconstruction from multiple 2-{D} forward-scan sonar views by space carving,'' \emph{IEEE J. of Oceanic Engineering (JOE)}, vol.~42, no.~3, pp. 574--589, Sep. 2016.

\bibitem{feng2024differentiable}
Y.~Feng, W.~Lu, H.~Gao, B.~Nie, K.~Lin, and L.~Hu, ``Differentiable space carving for 3{D} reconstruction using imaging sonar,'' \emph{IEEE Robotics and Automation Letters (RAL)}, vol.~9, no.~11, pp. 10\,065--10\,072, Sep. 2024.

\bibitem{teixeira2016underwater}
P.~V. Teixeira, M.~Kaess, F.~S. Hover, and J.~J. Leonard, ``Underwater inspection using sonar-based volumetric submaps,'' in \emph{Proc. IEEE/RSJ Intl. Conf. on Intelligent Robots and Systems (IROS)}, Daejeon, South Korea, Oct. 2016, pp. 4288--4295.

\bibitem{aykin20153}
M.~D. Aykin and S.~Negahdaripour, ``On 3-{D} target reconstruction from multiple 2-{D} forward-scan sonar views,'' in \emph{Proc. IEEE/MTS OCEANS Conf. and Exhibition}, Genova, Italy, May 2015, pp. 1--10.

\bibitem{westman2020theory}
E.~Westman, I.~Gkioulekas, and M.~Kaess, ``A theory of {F}ermat paths for 3{D} imaging sonar reconstruction,'' in \emph{Proc. IEEE/RSJ Intl. Conf. on Intelligent Robots and Systems (IROS)}, Las Vegas, NV, USA, Oct. 2020, pp. 5082--5088.

\bibitem{westman2019wide}
E.~Westman and M.~Kaess, ``Wide aperture imaging sonar reconstruction using generative models,'' in \emph{Proc. IEEE/RSJ Intl. Conf. on Intelligent Robots and Systems (IROS)}, Macau, China, Nov. 2019, pp. 8067--8074.

\bibitem{wang20183d}
Y.~Wang, Y.~Ji, H.~Woo, Y.~Tamura, A.~Yamashita, and A.~Hajime, ``3{D} occupancy mapping framework based on acoustic camera in underwater environment,'' \emph{IFAC-PapersOnLine}, vol.~51, no.~22, pp. 324--330, Dec. 2018.

\bibitem{wang2019three}
Y.~Wang, Y.~Ji, H.~Woo, Y.~Tamura, A.~Yamashita, and H.~Asama, ``Three-dimensional underwater environment reconstruction with graph optimization using acoustic camera,'' in \emph{In Proc. IEEE/SICE Intl. Symp. on System Integration (SII)}, Paris, France, Jan. 2019, pp. 28--33.

\bibitem{westman2020volumetric}
E.~Westman, I.~Gkioulekas, and M.~Kaess, ``A volumetric albedo framework for 3{D} imaging sonar reconstruction,'' in \emph{Proc. IEEE Intl. Conf. on Robotics and Automation (ICRA)}, Paris, France, May 2020, pp. 9645--9651.

\bibitem{debortoli2019elevatenet}
R.~DeBortoli, F.~Li, and G.~A. Hollinger, ``Elevate{N}et: A convolutional neural network for estimating the missing dimension in 2{D} underwater sonar images,'' in \emph{Proc. IEEE/RSJ Intl. Conf. on Intelligent Robots and Systems (IROS)}, Macau, China, Nov. 2019, pp. 8040--8047.

\bibitem{wang2021elevation}
Y.~Wang, Y.~Ji, D.~Liu, H.~Tsuchiya, A.~Yamashita, and H.~Asama, ``Elevation angle estimation in 2{D} acoustic images using pseudo front view,'' \emph{IEEE Robotics and Automation Letters (RAL)}, vol.~6, no.~2, pp. 1535--1542, Feb. 2021.

\bibitem{arnold2022spatial}
S.~Arnold and B.~Wehbe, ``Spatial acoustic projection for 3{D} imaging sonar reconstruction,'' in \emph{Proc. IEEE Intl. Conf. on Robotics and Automation (ICRA)}, Philadelphia, PA, USA, May 2022, pp. 3054--3060.

\bibitem{10.1145/3641519.3657446}
M.~Qadri, K.~Zhang, A.~Hinduja, M.~Kaess, A.~Pediredla, and C.~A. Metzler, ``A{ON}eu{S}: A neural rendering framework for acoustic-optical sensor fusion,'' in \emph{Proc. SIGGRAPH}, Denver, CO, USA, Jul. 2024, pp. 1--12.

\bibitem{10.1145/3592141}
A.~Reed, J.~Kim, T.~Blanford, A.~Pediredla, D.~Brown, and S.~Jayasuriya, ``Neural volumetric reconstruction for coherent synthetic aperture sonar,'' \emph{ACM Trans. on Graphics (TOG)}, vol.~42, no.~4, pp. 1--20, Jul. 2023.

\bibitem{10631294}
Y.~Xie, G.~Troni, N.~Bore, and J.~Folkesson, ``Bathymetric surveying with imaging sonar using neural volume rendering,'' \emph{IEEE Robotics and Automation Letters (RAL)}, vol.~9, no.~9, pp. 8146--8153, Sep. 2024.

\bibitem{10832111}
Y.~Xie, J.~Zhang, N.~Bore, and J.~Folkesson, ``Neu{RSS}: Enhancing {AUV} localization and bathymetric mapping with neural rendering for sidescan {SLAM},'' \emph{IEEE J. of Oceanic Engineering (JOE)}, pp. 1--10, Jan. 2025.

\bibitem{10685550}
Z.~Qu, O.~Vengurlekar, M.~Qadri, K.~Zhang, M.~Kaess, C.~Metzler, S.~Jayasuriya, and A.~Pediredla, ``Z-{S}plat: {Z}-axis gaussian splatting for camera-sonar fusion,'' \emph{IEEE Trans. on Pattern Analysis and Machine Intelligence (TPAMI)}, pp. 1--12, Sep. 2024.

\bibitem{kaess2012isam2}
M.~Kaess, H.~Johannsson, R.~Roberts, V.~Ila, J.~J. Leonard, and F.~Dellaert, ``i{SAM}2: Incremental smoothing and mapping using the bayes tree,'' \emph{Intl. J. of Robotics Research (IJRR)}, vol.~31, no.~2, pp. 216--235, May 2012.

\bibitem{dellaert2017factor}
F.~Dellaert and M.~Kaess, ``Factor graphs for robot perception,'' \emph{Foundations and Trends{\textregistered} in Robotics}, vol.~6, no. 1-2, pp. 1--139, 2017.

\bibitem{qadri2022incopt}
M.~Qadri, P.~Sodhi, J.~G. Mangelson, F.~Dellaert, and M.~Kaess, ``In{CO}pt: Incremental constrained optimization using the bayes tree,'' in \emph{Proc. IEEE/RSJ Intl. Conf. on Intelligent Robots and Systems (IROS)}, Kyoto, Japan, Oct. 2022, pp. 6381--6388.

\bibitem{qadri2024learning}
M.~Qadri, Z.~Manchester, and M.~Kaess, ``Learning covariances for estimation with constrained bilevel optimization,'' in \emph{Proc. IEEE Intl. Conf. on Robotics and Automation (ICRA)}, Yokohama, Japan, May 2024, pp. 15\,951--15\,957.

\bibitem{qadri2023learning}
M.~Qadri and M.~Kaess, ``Learning observation models with incremental non-differentiable graph optimizers in the loop for robotics state estimation,'' in \emph{Intl. Conf. on Machine Learning (ICML) 2023 Workshop on Differentiable Almost Everything: Differentiable Relaxations, Algorithms, Operators, and Simulators}, Jul. 2023.

\bibitem{Shin15Ocean}
Y.-S. Shin, Y.~Lee, H.-T. Choi, and A.~Kim, ``Bundle adjustment from sonar images and {SLAM} application for seafloor mapping,'' in \emph{Proc. IEEE/MTS OCEANS Conf. and Exhibition}, DC, USA, Oct. 2015, pp. 1--6.

\bibitem{KAZE}
P.~F. Alcantarilla, A.~Bartoli, and A.~J. Davison, ``{KAZE} features,'' in \emph{Proc. Eur. Conf. on Computer Vision (ECCV)}, Florence, Italy, Oct. 2012, pp. 214--227.

\bibitem{Huang15iros}
T.~A. Huang and M.~Kaess, ``Towards acoustic structure from motion for imaging sonar,'' in \emph{Proc. IEEE/RSJ Intl. Conf. on Intelligent Robots and Systems (IROS)}, Hamburg, Germany, Sep. 2015, pp. 758--765.

\bibitem{westman2019degeneracy}
E.~Westman and M.~Kaess, ``Degeneracy-aware imaging sonar simultaneous localization and mapping,'' \emph{IEEE J. of Oceanic Engineering (JOE)}, vol.~45, no.~4, pp. 1280--1294, Oct. 2020.

\bibitem{Loi24Ocean}
N.~Loi, Y.~Z. Tan, E.~W. Goh, and M.~H. Ang, ``Sonar {SLAM} in structured underwater environments,'' in \emph{Proc. IEEE/MTS OCEANS Conf. and Exhibition}, Singapore, Apr. 2024, pp. 1--10.

\bibitem{Xu24ICRA}
S.~Xu, K.~Zhang, Z.~Hong, Y.~Liu, and S.~Wang, ``{DISO}: Direct imaging sonar odometry,'' in \emph{Proc. IEEE Intl. Conf. on Robotics and Automation (ICRA)}, Yokohama, Japan, May 2024, pp. 8573--8579.

\bibitem{wang2022nerfneuralradiancefields}
\BIBentryALTinterwordspacing
Z.~Wang, S.~Wu, W.~Xie, M.~Chen, and V.~A. Prisacariu, ``{NeRF}{-}{-}: Neural radiance fields without known camera parameters,'' 2022. [Online]. Available: \url{https://arxiv.org/abs/2102.07064}
\BIBentrySTDinterwordspacing

\bibitem{Lin2021BARFBN}
C.-H. Lin, W.-C. Ma, A.~Torralba, and S.~Lucey, ``{BARF}: Bundle-adjusting neural radiance fields,'' in \emph{Proc. Intl. Conf. on Computer Vision (ICCV)}, Montreal, Canada, Oct. 2021, pp. 5721--5731.

\bibitem{Bian2022NoPeNeRFON}
W.~Bian, Z.~Wang, K.~Li, J.~Bian, and V.~A. Prisacariu, ``{NoPe-NeRF}: Optimising neural radiance field with no pose prior,'' in \emph{Proc. IEEE Intl. Conf. on Computer Vision and Pattern Recognition (CVPR)}, Vancouver, Canada, Jun. 2023, pp. 4160--4169.

\bibitem{Chng2022GaussianAN}
S.-F. Chng, S.~Ramasinghe, J.~Sherrah, and S.~Lucey, ``Gaussian activated neural radiance fields for high fidelity reconstruction and pose estimation,'' in \emph{Proc. Eur. Conf. on Computer Vision (ECCV)}, Tel Aviv, Israel, Oct. 2022, pp. 264--280.

\bibitem{Park2023CamPCP}
K.~Park, P.~Henzler, B.~Mildenhall, J.~T. Barron, and R.~Martin-Brualla, ``Cam{P}: Camera preconditioning for neural radiance fields,'' \emph{ACM Trans. on Graphics (TOG)}, vol.~42, no.~6, pp. 1--11, Dec. 2023.

\bibitem{yariv2020multiview}
L.~Yariv, Y.~Kasten, D.~Moran, M.~Galun, M.~Atzmon, B.~Ronen, and Y.~Lipman, ``Multiview neural surface reconstruction by disentangling geometry and appearance,'' in \emph{Proc. Advances in Neural Information Processing Systems (NeurIPS)}, vol.~33, Vancouver, Canada, Dec. 2020, pp. 2492--2502.

\bibitem{gropp2020implicit}
\BIBentryALTinterwordspacing
A.~Gropp, L.~Yariv, N.~Haim, M.~Atzmon, and Y.~Lipman, ``Implicit geometric regularization for learning shapes,'' 2020. [Online]. Available: \url{https://arxiv.org/abs/2002.10099}
\BIBentrySTDinterwordspacing

\bibitem{Westman18tr}
E.~Westman and M.~Kaess, ``Underwater {A}pril{T}ag {SLAM} and calibration for high precision robot localization,'' Robotics Institute, Carnegie Mellon University, Tech. Rep. CMU-RI-TR-18-43, Oct. 2018.

\bibitem{HoloOceanDocs}
{Field Robotic Systems Lab (FRoStLab), Brigham Young University}, ``Hovering{AUV} - {HoloOcean} 1.0.0 documentation,'' \url{https://byu-holoocean.github.io/holoocean-docs/UE5.3_Prerelease/agents/hovering-auv-agent.html}.

\bibitem{Potokar24joe}
E.~Potokar, K.~Lay, K.~Norman, D.~Benham, S.~Ashford, R.~Peirce, T.~Neilsen, M.~Kaess, and J.~Mangelson, ``{HoloOcean}: A full-featured marine robotics simulator for perception and autonomy,'' \emph{IEEE J. of Oceanic Engineering (JOE)}, vol.~49, no.~4, pp. 1322--1336, Oct. 2024.

\bibitem{BluefinHAUV}
{General Dynamics Mission Systems}, ``Bluefin {HAUV},'' \url{https://gdmissionsystems.com/products/underwater-vehicles/bluefin-hauv}.

\bibitem{DIDSON}
{Sound Metrics}, ``{DIDSON} 300m: {OBSERVE AND CONQUER},'' \url{https://blueprintsubsea.com/oculus/oculus-m-series}.

\bibitem{instantngp}
T.~M\"{u}ller, A.~Evans, C.~Schied, and A.~Keller, ``Instant neural graphics primitives with a multiresolution hash encoding,'' \emph{ACM Trans. on Graphics (TOG)}, vol.~41, no.~4, pp. 1--15, Jul. 2022.

\end{thebibliography}

\end{document}